\documentclass[12pt]{article}
\usepackage{epsfig}
\begin{document}
\title{Macroscopic properties of nuclei according to the relativistic mean 
       field theory\thanks{The work was partially sponsored by the State 
       Committee for Scientific Research under contract No. 2~P03B 115 19 
       and the collaboration between IN2P3 and Polish laboratories nr 99-95}}

\author{ Bo\.zena Nerlo-Pomorska, Katarzyna Mazurek           \\ 
        {\it Department of Theoretical Physics, Institute of Physics,}  \\ 
        {\it Maria Curie-Sk\l{}odowska University, 20-031 Lublin, Poland }} 
        
\date{}

\maketitle

\begin{abstract}
Self-consistent calculations within the relativistic mean field theory  (RMFT)
were performed for 150 spherical even-even nuclei.  The macroscopic part of the
binding energy was evaluated by subtracting  the Strutinsky shell corrections
from the RMFT energy. The parameters of  a liquid-drop (LD), like mass formula which
approximates the RMFT results, were  determined. The mass and isospin dependence
of the RMFT mean-square radii constant  for the neutron, proton, charge and
total density distributions were estimated. The RMFT liquid-drop parameters and
the radii constants are compared with similar results obtained with the
Hartree-Fock-Bogoliubov calculations with the Gogny force (HFB+Gogny)
and phenomenological models.
\end{abstract}

\section{Introduction}

Self-consistent Hartree-Fock (HF) calculations with effective nucleon-nucleon forces
of the Gogny \cite{berg} and  Skyrme \cite{chab} type, or within the relativistic
mean field theory (RMFT) \cite{rin} are nowadays able to describe many features
of nuclei. The theoretical results agree with the experimental data
and the masses, charge and neutron radii, electric multipole moments or energies of the
lowest excited  states   are well reproduced even for nuclei beyond the
stability line. It is probable that the presently accessible nuclei with 
the large neutron excess
demand some revision of the parameters used in the traditional models,
which have been adjusted to the smaller amount of data around the $\beta$ stable
nuclei. 

It is also interesting to compare the self-consistent
 prescriptions with  other simpler models and see how they work for the
nuclei close to the proton or neutron drip lines. The macroscopic-microscopic
method with the liquid-drop (or droplet) model, using the Strutinsky shell
correction and various kinds of the  single-particle average potentials is of special
interest because of its simplicity. Is it possible to extract the shell effects
from the self-consistent energy and obtain an estimate of the macroscopic energy
hidden in these models? This was already done successfully for the Skyrme [4] and Gogny
[5] forces and now we would like to apply a similar method for the relativistic
mean field (RMF). Nevertheless, one has to remember that the weak binding effects
at drip lines are of purely quantal origin  and the application of macroscopic-microscopic
method could be questionable there.

In Section II a short overview of the RMFT equations and parameters is given
and the prescription for shell correction [6] is recalled. Moreover the
  liquid-drop
formulae for the macroscopic energy and the root mean square (r.m.s.) radii of
the proton, neutron and charge distributions are mentioned. In Section III the
macroscopic part of the RMFT energy is approximated by a liquid-drop like
formula. The liquid-drop parameters corresponding to the macroscopic part
of the RMFT binding energy are compared with  those of
other theoretical and phenomenological models. In Section IV the RMFT root
 mean-square radii of 150 even-even spherical nuclei are  approximated by the
  isospin-dependent
formulae and compared with other experimental and theoretical estimates.
The ratio of proton to neutron radius is of special interest due to the
lack of experimental data for neutron radii. One can use this ratio to predict
the neutron r.m.s. radius of a nucleus when its charge radius is known. At the
end of the paper the conclusions are drawn and further investigations
 proposed.

\section{Theory}

The single-particle level scheme obtained within the self-consistent RMFT
calculation is used to evaluate the shell correction  ($E_{\rm shell}$) 
to the binding energy:
\begin{equation}
 E_{\rm shell} = \sum_{occ} 2e_\nu - \tilde E \,\,,
\end{equation}
where the sum runs over all occupied levels.
All the single-particle levels up to the cut-off energy lying 15~MeV
 above the Fermi surface, 
 are used to obtain the  smoothed energy from the Strutinsky integral.
We haven't apply the newer prescription  for the shell correction
proposed in Ref. \cite{krup}  to avoid the single-particle continuum 
effect because  we wished to compare our results with
those of reference \cite{kleb} obtained with the classical Strutinsky 
prescription.  Nevertheless, the single-particle levels
scheme for each nucleus, especially for the neutron-rich ones was carefully
 checked in
order to take into account the proper number of single-particle states around
Fermi surface. It was not our aim to estimate the position of the drip lines,
 but to obtain 
the average dependence of binding energy on $A, Z$ number. For the detailed calculation of the 
binding energies of the nuclei close to the proton or neutron drip lines
 the use of the prescription of Ref. \cite{krup,vert} would be necessary.
  The Strutinsky smooth 
 energy $\tilde E$ is equal to
\begin{equation} 
 \tilde E = 2\int^\lambda_{-\infty} e\bar\rho(e)de \,\,.
\end{equation}
The average levels  density $\bar \rho(e)$  was obtained by the smoothing 
of the single-particle levels density $\rho(e)= \sum_\nu \delta(e-e_\nu)$ with 
the Gauss function multiplied by the 6th order correction polynomial~$f$
\begin{equation}
\bar \rho(e) = {1\over\gamma\sqrt{\pi}} \int\limits^{+\infty}_{-\infty} 
    \rho(e') \,e^{-\left({e-e'\over\gamma}\right)^2}\,f\left({e-e'\over\gamma}
    \right) de'\,\,.
\end{equation}
The width parameter of the Gauss function $\gamma = 1.2 \hbar\omega,~~{\rm with}
~~~\hbar\omega = 40 A^{-1/3}~{\rm MeV}\,\,,$
corresponds to the average position of the Strutinsky plateau 
in the shell corrections for the chosen sample of 150 spherical even-even nuclei.
The average single-particle levels density obtained with the  RMFT is close to the results
obtained in Ref. \cite{kleb} for the Gogny force.                                                                  

The macroscopic part of the binding energy is equal to the difference between 
the self-consistently calculated RMFT energy ($E_{\rm RMFT}$) without pairing
 interaction and 
the total (neutron and proton) shell correction 
\begin{equation}
 E^{\rm RMFT}_{\rm macr} = E_{\rm RMFT} - E^n_{\rm shell} - E^p_{\rm shell} \,\,.
\label{macr}
\end{equation}
These quantities, evaluated for several nuclei with mass numbers $A$ and 
isospins $I=(N-Z)/A$
are approximated by the liquid-drop formula of Myers-\'Swi\c atecki type \cite{mye}
\begin{equation}
 E_{\rm macr} = -b_{\rm vol} (1 - \kappa_{\rm vol} \,I^2) A +
 b_{\rm surf} (1 - \kappa_{\rm surf}\, I^2) A^{2/3} 
 + b_{\rm Coul} Z^2 A^{-1/3} - C_4 Z^2/A
\label{msld}
\end{equation}
where $b_{\rm Coul}$ is connected with the charge radius parameter  
$r^{\rm ch}_0$ by $ b_{\rm Coul} = {3\over 5} e^2 /r^{\rm ch}_0\,.$

The nucleon densities $\rho_n(\vec r)$ and $\rho_p(\vec r)$ obtained within 
the RMFT + BCS model could be used to evaluate the mean-square radii of the 
neutron or proton distributions. Here the question of the validity of that
 approach for the nuclei near drip line rises again, but the influence of pairing
forces on the nuclear radius calculated in various models ( HFB + Gogny, LD + Woods
Saxon ) is similar and does not interfere with the isotopic shifts, which 
have been measured
\begin{equation}
\langle r^2\rangle_q = \int \rho_q(\vec r)\, {\vec r}\,^2\, d V ~/~
         \int \rho_q(\vec r) \, d V \,\,,~~~~~~~q=\{n,p\} \,\,.
\label{msr}
\end{equation}
Knowing the mean square radius $\langle r^2\rangle$ one can define an
equivalent spherical sharp radius $R$ using the following relation
$\langle r^2\rangle =  {3\over 5} R^2\,\,,$
which arises directly from Eq. (\ref{msr}) for the uniform density distribution 
$\rho_q = {{\cal N}_q\over 4/3\pi R^3}$, with 
$ {\cal N}_q=\{N,Z\}\,\,$
and the volume conservation condition. In a rough estimate one usually
assumes that $ R = r_0 A^{1/3}\,\,,$
and takes the radius constant $r_0 = 1.2$~fm.
 However this formula turns out to be too
approximate and it was proved in Ref. \cite{nerlo} that a similar formula but 
using an isospin dependent radius constant, described the experimental
data in a more satisfactory way. We have shown that the measured or calculated mean-square radii within
the RMFT \cite{war}  or HFB+Gogny \cite{kleb} models could be accurately
reproduced when the radius constant has the following 
form
\begin{equation}
 r_0 = r_{00} (1 + \alpha I + \kappa/A)\,,
\label{r0}
\end{equation}
where $r_{00}$, $\alpha$, and $\kappa$ are free adjustable
parameters.
The ratio of the proton to neutron root mean-square radii could 
be described by a formula similar to the one given above and could be used
 to predict
the radius of the neutron distribution when the charge radius is measured
\cite{war,kleb}. One has to note that this ratio does not depend on 
deformation in a first approximation since the density distributions of neutrons
and protons are close to each other also for deformed nuclei.

\section{Binding energies}

The RMFT calculations with  the NL3 set of parameters were performed  for 150
even-even nuclei between the proton and neutron drip lines which have, 
according to Ref.~\cite{moel} a quadrupole moment almost equal to zero. They 
are: $^{38-50}$Ca, $^{82-90}$Sr, $^{96-140}$Sn, $^{80-84}$Sm, $^{162-220}$Pb
isotopes, $N=50$ with $A \in (86, 92)$, $N=82$ with $A \in (122, 164)$,  and $N=126$ with
$A \in (174, 224)$ isotones and 30 other spherical nuclei along  the $\beta$
stability line. This choice of nuclei had already been used to estimate  the shell
effects by the HF method with the Gogny force \cite{kleb}. This set of 
representative spherical nuclei is larger than the sample of 30 deformed nuclei
taken for the radii calculation within the RMFT in Ref. \cite{war}.

We have used $N_0 = 20$ shells and the oscillator length constant $b = 2.4$~MeV of the harmonic
oscillator as the basis when solving the self-consistent RMFT equations for
fermions. At first the calculations were performed without taking into account
the pairing residual interaction in order to evaluate the Strutinsky  shell
corrections, and then the experimental proton and neutron pairing  energy gaps
$\Delta_p$, $\Delta_n$ were used to evaluate the r.m.s. radii and the potential
energies in the RMFT + BCS model. This simplified  way of pairing correlation
 inclusion does not
influence the values of radii significantly even for the nuclei near the
 drip lines.

\subsection{Liquid-drop parameters}

 Fig. 1 shows the RMFT (solid lines) shell corrections in comparison
with the results of Ref. [5] (dashed lines) obtained for the Gogny force. In the
first panel  of the multiplot one can observe the dependence on $A$ of the total shell
correction $E^{tot}_{\rm shell}$ for six groups of Ca-Th isotopes, in the middle 
$E^{tot}_{\rm shell}$ for three groups of $N = 50, 82, 126$ isotones and
in the r.h.s. panel for $\beta$ stable isotopes.

The shell corrections obtained in both theoretical models are similar. They
exhibit  minima for the same magic numbers of one kind of nucleons and differ from each
other by no more than a few MeV. 
The RMFT estimates of the macroscopic part of the binding energy obtained  by
subtracting the total shell correction from the self-consistent RMFT energy  
(Eq. \ref{macr}) for the above set of nuclei were fitted by the liquid-drop 
formula (\ref{msld}), and the following set of parameters was obtained
\begin{equation}
{E^{\rm RMFT}_{\rm macr}\over {\rm MeV}} = -15.19  (1 - 1.66 I^2)A + 
 16.81(1 - 1.21 I^2)A^{2/3} + 0.68{ Z^2\over A^{1/3}}  - 1.3{Z^2\over A}\,\,.
\end{equation}
The r.m.s. deviation of the fit was  equal to  1.97 MeV. In the Table~I these
 RMFT estimates
of the LD parameters are compared with the traditional (MS-1967)
Myers-\'Swi\c{a}tecki liquid-drop formula \cite{mye}. What is more Table~I
 shows the modern
phenomenological   set (MS-2002) \cite{pomor} fitted to presently available
experimental masses  \cite{ant}  when using the microscopic
(shell+pairing+deformation) energy corrections from Ref. \cite{moel}. In the 
last column of Table~I are given the results 
 obtained in \cite{kleb} within the Hartree-Fock calculation
with the Gogny D1S force \cite{berg} which turned ut to be similar to these of 
RMFT.

During the last 35 years, as seen in Table~I, the liquid-drop
parameters  reproducing the experimental data have not changed very much.  The
macroscopic part of the binding energies obtained with the Gogny force 
\cite{kleb} is described by the set of the LD parameters which approximates to the
newest fit (MS-2002) of the LD parameters adjusted to the presently known experimental
 masses \cite{ant}. The
results obtained within the RMFT give  smaller values of the volume and
 surface energies, while the  charge radius constant corresponding
to the Coulomb energy is equal to
1.264~fm  and is substantially larger than its present phenomenological value
(1.191 fm). By contrast, the RMFT estimate of the $C_4$ parameter, which is
responsible for the charge diffuseness effect, is much closer to its
phenomenological value compared with the  Gogny's one. The volume and surface
 dependence on 
isospin is weaker in the RMFT than the experimental
one. The Gogny force gives a slightly stronger dependence of both
energies than the phenomenological (MS-2002) one.

We can compare the three models in Fig. 2. The results of MS-2002 liquid-drop
 model and Gogny
 are subtracted  from the macroscopic
energies of the RMFT and
 shown for all the groups of
nuclei in dependence on $A$. Since the RMFT macroscopic energy (solid lines) is the
smallest,  it gives the largest binding. The Gogny results 
are closer to the RMFT ones than to the phenomenological (MS-2002)  binding
 energy.

The differences between the binding energies obtained with these three models
 reach even  -30~MeV for isotope and 
isotone chains while
for $\beta$ stable isotopes they stay within -20~MeV. This is
understandable because the NL3 set of parameters
of the RMFT was fitted for nuclei close to the $\beta$ stability line.
The binding energies obtained with the Gogny force are closer  
to the liquid-drop estimatesthan these 
of RMFT. The isospin dependence of the binding energies 
is not well reproduced by either of the both models. 

\section{Mean-square radii}

It is a known fact 
that the pairing correlations influence the density distribution in nuclei. 
Therefore in order to evaluate the neutron and proton mean-square radii within the 
RMFT, we have to include the pairing forces. This was done in a simplified way by inserting into
the BCS equations the experimental proton and neutron energy gap between the
ground state and the first excited two-quasiparticle state of even-even nuclei.
The pairing energy gaps are extracted from the experimental binding energies
\cite{audi} with the help of a three parameter formula proposed in Ref.
\cite{sat}
\begin{equation}
 \Delta_q = {\pi_{{\cal N}_q}\over 2}[(B({\cal N}_q - 1) - 2B({\cal N}_q) + 
                 B({\cal N}_q+1)]\,\,,~~~~~~~q=\{n,p\} \,\,,
\end{equation}
where $\pi_{{\cal N}_q} = (-1)^{{\cal N}_q}$ and ${\cal N}_q$ denotes nucleon number 
$N$ for neutrons, $Z$ for protons. When the BCS equations are solved the
pairing correlations are added to the self-consistent mean field.

The resulting r.m.s. radii for neutron and charge distributions as well as the
ratio of proton to neutron radii are plotted in Figs. 3~-~5 for the three
groups of  isotopes, isotones and $\beta$ stable nuclei. The RMFT radii can be
easily reproduced by the isospin dependent formula (\ref{r0}), which corresponds
to the sharp density distribution. The RMFT radius constants  fitted for the 150
spherical nuclei are:\\
for neutrons
\begin{equation}
 r^n_0 = 1.17 (1 + 0.27 I + 3.38/A)~{\rm fm} \,\,,
\label{r0n}
\end{equation}
for protons
\begin{equation}
 r^p_0 = 1.22 (1 - 0.15 I + 1.51/A)~{\rm fm} \,\,,
\label{r0p}
\end{equation}
and for charge distribution
\begin{equation}
r^{\rm ch}_0 = 1.23 (1 - 0.15 I + 2.47/A)~{\rm fm} \,\,.
\label{r0ch}
\end{equation}
The r.m.s. deviation   of each fit was smaller than 0.01 fm. The estimates
(\ref{r0n}-\ref{r0ch}) are very close to those obtained in Ref. \cite{war}
for the smaller sample (30) of deformed nuclei. This means that the
deformation dependent function renormalizing the distributions to the sphere
was properly chosen in  Ref.~\cite{war}, and that the formulae
(\ref{r0n}-\ref{r0ch}) adequately describe the radii constants , not only for the
spherical but also for the deformed nuclei.

This is also the case with the proton to neutron ratio
\begin{equation}
 {r_p\over r_n} = 1.04 (1 - 0.38I - 1.52/A)\,,
\label{rporn}
\end{equation}
which can be used to estimate the neutron radii with the help of the measured 
charge radius 
\begin{equation}
 r_n = {\sqrt{r^2_{\rm ch} - 0.64 {\rm fm^2}}\over 1.04(1 - 0.38I - 1.52/A)}
\end{equation}
producing  good agreement (with a slight tendency to overestimate) with the 14
experimentally known neutron radii of Ref.~\cite{fri}. In contrast, a similar ratio obtained
with the Gogny force in \cite{kleb} gives a slightly smaller neutron radius.
Both groups of neutron radii for the 14 experimentally known data can be seen
in Fig.~3 in dependence on the reduced isospin I.

In Fig.~4 the differences between the
charge radii predicted by the RMFT and the Gogny model (solid lines)  are compared with the
experimental data  \cite{fri} from which also the Gogny radii are subtracted
 (crosses). One can see that the agreement of
the RMFT results for the charge radii with the experimental data is even
 slightly better 
 than that  of Ref. \cite{kleb} obtained with the Gogny force.

In Fig. 5 the proton to neutron radius ratio obtained in the RMFT (solid lines)
and with the Gogny force (dashed lines) is compared with the experimental data
(crosses) \cite{fri,bat} for the three groups of isotopes, isotones and $\beta$ 
stable nuclei.

The parameters of formulae (\ref{r0n}-\ref{r0ch}) obtained for 
various theoretical models are compared in Table II with the ones fitted 
 to 
the experimental data \cite{fri,bat} and in Ref. \cite{nerlo} for charge radii.

Both  self-consistent theoretical models give similar estimates of the $r_{00}$
parameter of neutron, proton, and charge radii . The isospin
dependence of the r.m.s.  radii is slightly different in the two models.
The $\kappa/A$ term, important for the light nuclei, shows some
differences as well. 

\section{Conclusions}

The following conclusions can be drawn from our investigation:
\begin{itemize}
\item[1.] The shell corrections obtained in the RMFT with the NL3 set of
          parameters and within the Hartree-Fock mean field calculation with 
          the Gogny D1S force are similar.
\item[2.] The volume and surface parts of the binding energy in the RMFT are 
          smaller than the corresponding energies obtained with the Gogny model 
          \cite{kleb} as well as  than those of the liquid-drop model
	  fitted to the experimental masses 
          \cite{mye,pomor}.
\item[3.] The isospin dependence of the volume and surface term obtained within
          the RMFT is too small in comparison with the phenomenological 
	  liquid-drop model.
\item[4.] The mean-square radii of the proton, neutron and charge distributions
          are similar in the Gogny and RMFT models.
\item[5.] The RMFT ratio of proton to neutron radii, used to predict the
          neutron radii when the charge radius is known, gives the estimates
          within experimental error bars for all 14 experimentally known neutron
          radii.
\end{itemize}

Similar effects for deformed nuclei with various sets of RMFT parameters will be investigated soon.	  

\bigskip
\noindent
{\bf Acknowledgments}

We would like to thank professors Klaus Dietrich and Peter Ring for fruitful 
discussions and the warm hospitality during our stay at the Technische 
Universit\"at M\"unchen. The help  of  Krzysztof Pomorski  
 in formulating the manuscript is also appreciated.

\newpage

\begin{table}[h]
\caption[TT]{The macroscopic energy parameters.}
\label{tab1}
\begin{center}
\begin{tabular}{|c|c|c|c|c|c|}
\hline
parameter           & unit & MS-1967 & MS-2002&  RMFT  &  Gogny  \\
\hline
$b_{\rm vol}$       & MeV   & 15.667 & 15.848 & 15.185 & 15.649  \\
$\kappa_{\rm vol}$  & --    &  1.790 &  1.848 &  1.657 &  1.916  \\
$b_{\rm surf}$      & MeV   & 18.560 & 19.386 & 16.811 & 18.928  \\
$\kappa_{\rm surf}$ & --    &  1.790 &  1.983 &  1.209 &  2.108  \\
$r^{\rm ch}_0$      & fm    &  1.205 &  1.190 &  1.264 &  1.188  \\
$C_4$               & MeV   &  1.211 &  1.200 &  1.299 &  2.015  \\
\hline
\end{tabular}
\end{center}
\end{table}

%\newpage

\begin{table}[h]
\caption[TT]{The radii parameters.}
\label{tab2}
\begin{center}
\begin{tabular}{|c|c|c|c|c|}
\hline
& Phen. & Gogny (150 sph.n.)& RMFT (30 def.n) & RMFT (150 sph.n.) \\
\hline 
\underline{neutrons} & \cite{bat}&\cite{kleb}&\cite{war}&\\
$r_{00}$ & 1.17 & 1.17 & 1.17 & 1.17 \\
$\alpha$ & 0.16 & 0.12 & 0.25 & 0.27 \\
$\kappa$ & 3.85 & 3.29 & 2.81 & 3.38 \\
\hline
\underline{protons} &\cite{fri}&&& \\
$r_{00}$ & 1.22 & 1.21 & 1.24 & 1.22 \\
$\alpha$ & --0.17 & --0.14 & --0.16 & --0.15 \\
$\kappa$ & 1.78 & 1.83 & 0.65 & 1.51 \\
\hline
\underline{charge} &\cite{fri}(\cite{nerlo})&&& \\
$r_{00}$ & 1.24(1.25) & 1.22 & 1.24 & 1.23 \\
$\alpha$ & --0.19(--0.25) & --0.15 & --0.15 & --0.15 \\
$\kappa$ & 1.65(2.06) & 2.32 & 0.58 & 2.47 \\
\hline
\underline{ratio} &\cite{fri,bat}&&& \\
$r_{00}$ & 1.03 & 1.04 & 1.05 & 1.04 \\
$\alpha$ & --0.36 & --0.27 & --0.36 & --0.38 \\
$\kappa$ & --1.33 & --1.12 & --3.15 & --1.52 \\
\hline
\end{tabular}
\end{center}
\end{table}

\newpage

\begin{figure}
\begin{center} \epsfxsize=150mm \epsfbox{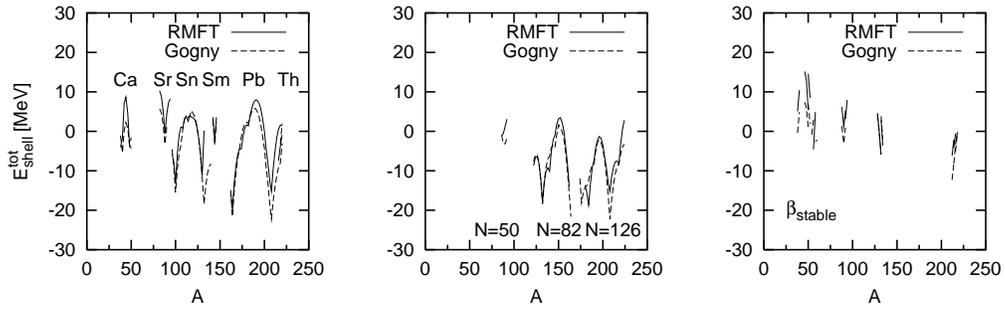} \end{center}

\caption{ The total shell corrections obtained within the RMFT (solid lines) and
with the Gogny force (dashed lines) in dependence on the mass number $A$.
The three parts of multiplot show the shell corrections of Ca-Th
isotopes, for the $N=50, 82, 126$ isotones and for the $\beta$ stable nuclei respectively. }

\end{figure}                                                                    

\newpage

\begin{figure}
\begin{center} \epsfxsize=150mm \epsfbox{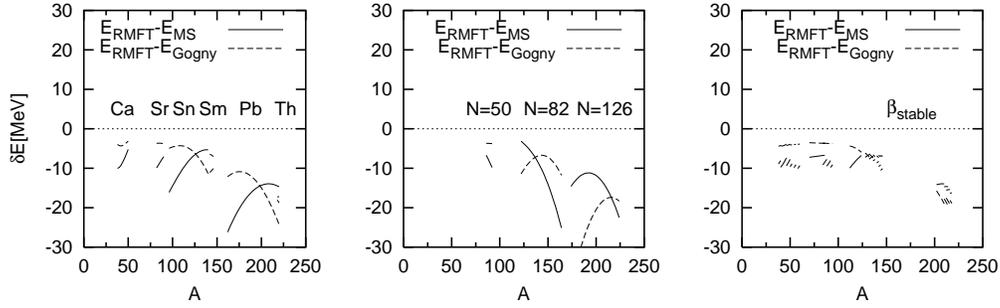} \end{center}

\caption{The comparison of macroscopic energies obtained within the RMFT 
 and with the Gogny \protect\cite{kleb}  force and the 
 liquid-drop energies \protect\cite{pomor} in dependence on $A$. The 
differences between the  RMFT macroscopic parts of the binding energies (solid
lines) and  the phenomenological (MS-2002) \protect\cite{pomor} (solid lines)
estimates are compared with the corresponding differences between RMFT and the Gogny
\protect\cite{kleb} macroscopic energies (dashed lines) for the  isotopes (l.h.s.) 
isotones (middle) and  $\beta$ stable nuclei(r.h.s.). }

\end{figure}                                                                    

\newpage

\begin{figure}
\begin{center} \epsfxsize=130mm \epsfbox{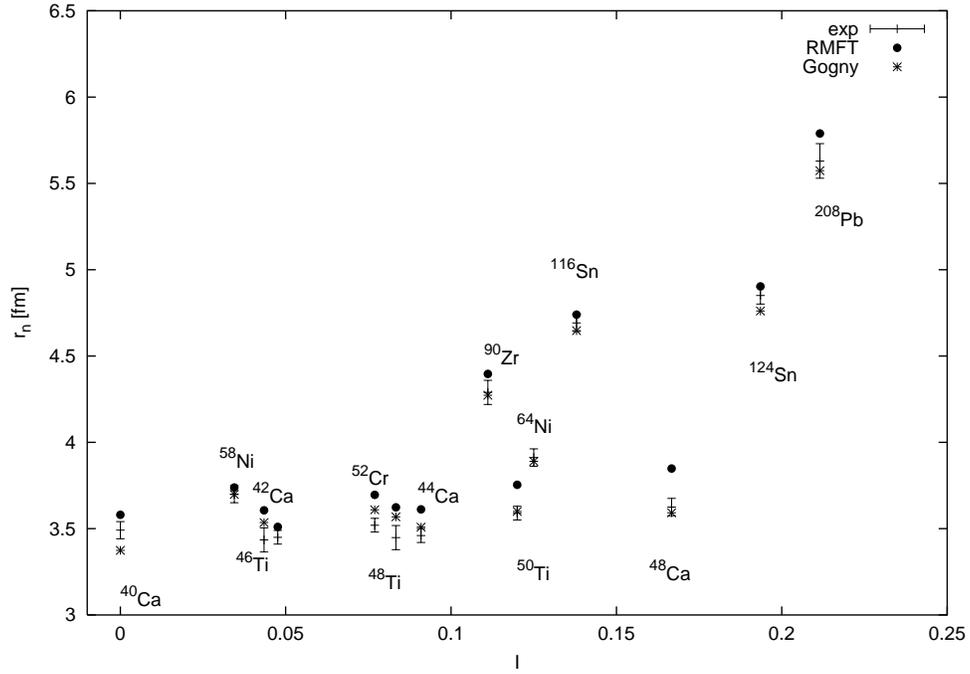} \end{center}

\caption{The 14 experimentally known neutron radii from Ref.
\protect\cite{bat} (crosses) are compared with the RMFT predictions
(spheres) and the estimates done with the Gogny model (stars) in dependence on
 the reduced isospin $I=(N-Z/A)$. }

\end{figure}                                                                    

\newpage

\begin{figure}
\begin{center} \epsfxsize=130mm \epsfbox{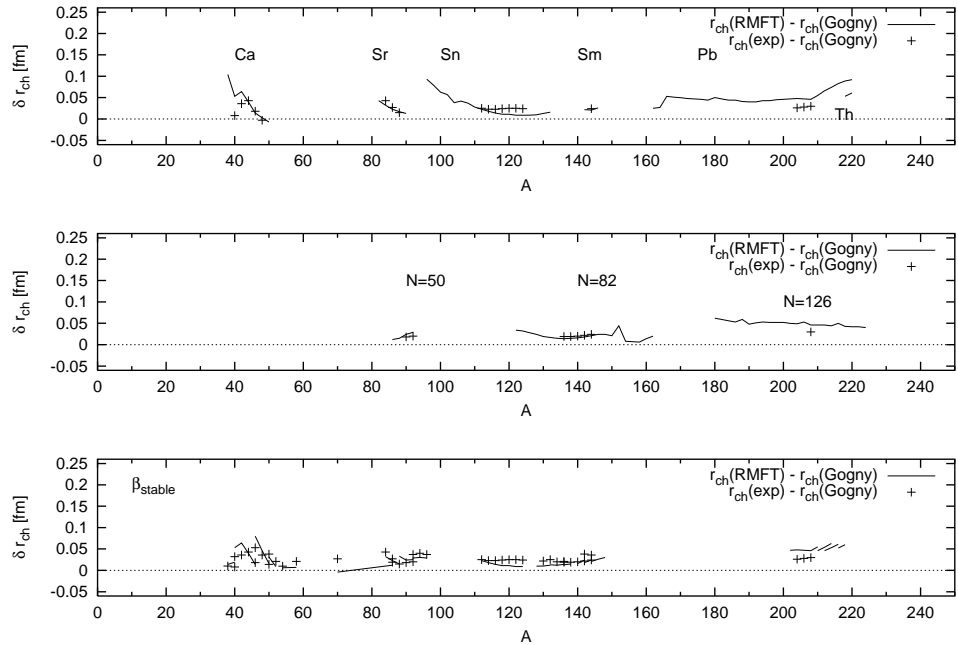} \end{center}

\caption{The charge radii  predicted by the RMFT 
related to the results of the Gogny model (solid lines) are compared with the experimental
data  \protect\cite{fri} from which the same reference of Gogny charge radii 
is also removed (crosses). The three panels correspond to the  isotopes (up) 
isotones (middle) and  $\beta$ stable nuclei (down). }

\end{figure}                                                                    

\newpage

\begin{figure}
\begin{center} \epsfxsize=150mm \epsfbox{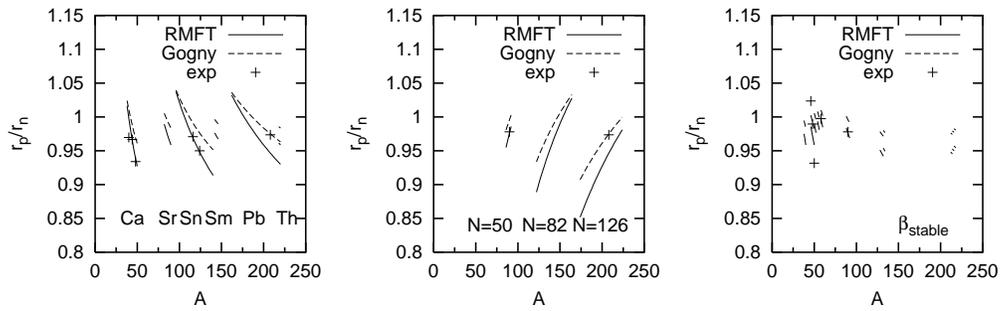} \end{center}

\caption{The proton to neutron radius ratio obtained in the RMFT (solid
line) and with the Gogny force (dashed lines) is compared with the similar ratio
obtained on the basis of the
experimental data (crosses) \protect\cite{fri,bat}. The three parts of multiplot
show the calculated root mean-square radii ratios for the three groups of 
isotopes (l.h.s.),
isotones (middle) and $\beta$ stable nuclei (r.h.s.) \protect\cite{kleb}.}

\end{figure}                                                                    

\end{document}